\documentclass[10pt,twocolumn,twoside]{IEEEtran}

\usepackage{cite}
\usepackage{graphicx}
\usepackage{subfigure}
\usepackage{algorithm2e}
\usepackage{amsmath}
\usepackage{balance}
\usepackage{multirow}
\usepackage{float}
\usepackage{stfloats}
\usepackage{flushend}
\usepackage{amssymb}
\usepackage[justification=centering]{caption}
\usepackage{mnsymbol}
\begin{document}
\title{An Interesting Property of LPCs \\
for Sonorant Vs Fricative  Discrimination}
\author{T. V. Ananthapadmanabha$^{1}$, \thanks{$^{1}$Voice and Speech Systems, Malleswaram, Bangalore, India 560003 (\mbox{e-mail: tva.blr@gmail.com}).} A. G. Ramakrishnan$^{2*}$, SMIEEE, \thanks{$^{2*}$Department of Electrical Engineering,
Indian Institute of Science, Bangalore,
India 560012 (\mbox{e-mail: ramkiag@ee.iisc.ernet.in}).} Pradeep Balachandran$^{3}$, MIEEE \thanks{$^{3}$Department of Electrical Engineering,
Indian Institute of Science, Bangalore,
India 560012 (\mbox{e-mail: pb@mile.ee.iisc.ernet.in}).}
}

\maketitle
\begin{abstract}
Linear prediction (LP) technique estimates an optimum all-pole filter of a given order for a frame of speech signal. The coefficients of the all-pole filter, 1/A(z) are referred to as LP coefficients (LPCs). The gain of the inverse of the all-pole filter, A(z) at z = 1, i.e, at frequency = 0, A(1) corresponds to the sum of LPCs, which has the property of being lower (higher) than a threshold for the sonorants (fricatives). When the inverse-tan of A(1), denoted as T(1), is used a feature and tested on the sonorant and fricative frames of the entire TIMIT database, an accuracy of 99.07\% is obtained. Hence, we refer to T(1) as sonorant-fricative discrimination index (SFDI).   This property has also been tested for its robustness for additive white noise and on the telephone quality speech of the NTIMIT database.  These results are comparable to, or in some respects, better than the state-of-the-art methods proposed for a similar task. Such a property may be used for segmenting a speech signal or for non-uniform frame-rate analysis.
\end{abstract}

\section{Introduction}

Integration of knowledge of phonetic classes into a statistical based ASR system  \cite{1,2,3} is known to supplement its performance.  This paper is concerned with the discriminability between two important phonetic classes, viz, sonorants and fricatives, from a continuous speech signal. The class of `sonorants' comprises vowels and voiced consonants excluding  voiced stops (b, d, g) - all voiced with the only exception being `hh' which is unvoiced. The class of ‘fricatives’ comprises unvoiced phones `s', `sh', `f’, `ch' and mixed voiced-unvoiced phones `z', `zh', `jh', `th', `dh' and 'v'. 

In the literature, this problem has been studied under the context of manner classification and landmark detection \cite{4,5,6} and also in the context of extraction of distinctive features \cite {7,8,9}. The problem of identifying sonorants Vs unvoiced fricatives may also be looked upon as a V/U classification problem, which has been extensively studied in the literature and we refer to some of them here \cite{10,11,12,13,14,15,16,17}.   

Several methods have been proposed for the identification of broad phonetic classes and/or their onsets from a speech signal. Liu \cite{5} has used the change of energy between two frames spaced 50 ms apart, over six sub-band signals, for detecting the onsets or landmarks of four broadly defined classes. Salomon et al \cite{6} have used a set of twelve temporal parameters to achieve manner classification. A team of researchers have used landmark based approach \cite{7} for feature extraction and experimented with different classifiers, such as SVMs, for identifying the distinctive features, which in turn may be used for manner classification. King and Taylor have used mel-frequency cepstral coefficients (MFCCs) and their derivatives (a 39-dimensional feature vector) to train a neural network to identify multi-dimensional distinctive features comprising broad manner classes \cite{8}. Juneja and Wilson combined MFCCs with certain temporal features and used an SVM classifier for the manner classification task \cite{9}. 

Most of the methods on V/U classification use the following temporal features \cite{11}: (i) the relative energy of a frame (which is typically low for unvoiced frames), (ii) the ratio of energies in the lowpass to highpass region (which is typically high for voiced segments), (iii) the number of zero-crossings per unit interval (which is typically high for the unvoiced segments), (iv) the value of normalized autocorrelation at one sample lag, which indirectly relates to the first reflection coefficient in linear
prediction (LP) analysis captures the gross spectral slope (typically lowpass for voiced and highpass for unvoiced), (v) periodicity detection (voiced sounds are periodic) and (vi) pitch prediction gain. Deng and O'Shaughnessy \cite{16} have used an unsupervised algorithm for V/U classification and tested the performance on the NTIMIT database.  Other features have also been considered. Alexandru Caruntu et al \cite{13} have used zero-crossing density, Teager energy and entropy measures. Dhananjaya and Yegnanarayana \cite{17} have used glottal activity detection, which in turn requires epoch extraction. Molla et al. \cite{18} have modeled the speech signal as a composite signal of intrinsic mode functions and extracted the trend of these functions. The trends are compared with thresholds obtained on a training data for V/U classification. 

In this paper, we demonstrate that the sum of linear prediction
coefficients (LPCs), a scalar measure, is useful in discriminating a
sonorant class from a fricative class in a speech signal. LP is a very successful speech analysis technique \cite{19,20,21}. According to the frequency domain interpretation \cite{21}, LP technique estimates an optimum all-pole digital filter, 1/A(z), that best approximates the short-time spectrum of a frame of speech signal. The reciprocal all-zero filter, A(z), called the digital inverse filter is given by
\begin{equation}
A(z)=1 + a_1z^{-1} + a_2z^{-2} + a_3z^{-3} +...+a_Mz^{-M}			
 \end{equation}
 
where ${a_1, a_2..., a_M}$ are the LPCs and M is the number of LPCs, which is a variable that can be set during the estimation of LPCs. Here $z^{-1}$ is the unit delay operator given by $z^{-1}=exp^{-j2\pi fT}$, where T is the sampling interval. Hence, the gain of the filter A(z) at $z = 1$ (or frequency = 0), i.e., A(1) is given by
\begin{equation}
A(z=1) = 1 + a_1 + a_2 + a_3 +...+a_M			
 \end{equation}
which corresponds to the sum of the LPCs.\\

In this paper, we report an interesting application of A(1) for segmentation of a speech signal into broad phonetic classes and test its effectiveness using the entire TIMIT database. 
\section{A(1) CONTOUR AND ITS CHARACTERISTIC}

Speech signal is divided into frames of 20 ms with two successive frames spaced by 5 ms. A Hanning window is applied on a frame of speech signal after removing the mean value for the frame. The autocorrelation method of LP technique \cite{20} is used as it assures the filter to be stable. The computed LPCs depend on the spectral shape, but are independent of the signal level.  The number of LPCs is chosen as $(F_s+2)$, where $F_s$ is the sampling frequency in kHz. LPCs are computed on the preemphasized and windowed speech signal. After obtaining the LPCs, the computed A(1) is assigned to the entire mid 5 ms of the speech frame. Thus a contour of A(1) appears like a staircase waveform. It has been found that the typical value of A(1) is greater than 13 for fricatives and less than 0.4 for sonorants.

The rationale for selecting A(1) as a feature arises as follows. Assuming the speech signal as the output of a quasi-stationary time-invariant filter (vocal tract) excited by an appropriate excitation, the intensity of a frame of speech signal is determined both by the strength of excitation (source intensity) and the filter gain. Given the speech signal, the source intensity can be obtained by an inverse filtering operation. It has been observed that the source intensity is higher for fricatives than for sonorants, which is exactly the opposite of what is observed for speech intensity. This may arise since there is a greater amount of acoustic loss for fricatives than for the resonant sounds of vowels. The filter gain itself is determined by two factors, namely, the frequency response and the filter gain at f = 0. Conventionally, a standard method of representing a filter is to set the filter gain at f = 0 to be unity. Hence we deduce that the filter gain at f = 0 must be influencing the source intensity. This led us to investigate A(1) as an acoustic feature for distinguishing sonorants from fricatives.  

An illustrative example: The utterance (sa2.wav) `Don't ask me to carry an oily rag like that' from the TIMIT database is analyzed. The speech wave and the computed A(1) values are shown in Fig. 1a for a part of the utterance. The hand labeled boundaries and the phone labels are also shown in the figure. Here A(1) is forced to zero for the silence frames, determined by a threshold on the energy. A(1) rises sharply at the onset of the fricative `s', reaches the maximum value of 28 and falls sharply at the end of the fricative segment. In Fig. 1a, we note that for sonorants, the maximum value of A(1) is low ($<$1.0). We make use of this property of A(1).  In this paper, we study the characteristic of A(1) for the 2-class problem of sonorants Vs fricatives. The maximum value of A(1) observed for fricatives is 119 and the minimum is 0.058 for sonorants. The large value of A(1) for fricatives dominates over the sonorants. The variation in A(1) within a fricative segment is not of interest as long as the value is above a threshold. Hence, we prefer to compress the range of A(1) using
\vspace{-0.05cm}
\begin{equation}
T(1) = tan^{-1}[A(1)]			
 \end{equation}
Such a compression of A(1) also helps in graphic visualization in comparison with a normalized signal waveform plot.  The upper bound for T(1) is $\pi/2$. During our investigation, we did not come across a single instance, where A(1) is negative.  Fig. 1b shows the plot of T(1), which compresses the range of A(1) and also swamps out the variations when A(1) is large. Henceforth, T(1) is termed as sonorant-fricative discriminant index (SFDI). 
 
For the stop segment (marked as `k' in Fig. 1b), T(1) reaches a maximum value close to 1.4 for a part of the segment. Stop bursts usually follow a silence or a low level voicing, which may be utilized for their detection \cite{22}.  Other phones also exhibit mixed characteristics for T(1). However, the detection of stop bursts and other phones is not a topic for this paper. 
 
%

      \begin{figure}
      \captionsetup{justification=raggedright,
      singlelinecheck=false
      }
      \centering{
      \subfigure[]{
      \includegraphics[width=3.5in,height=2in]{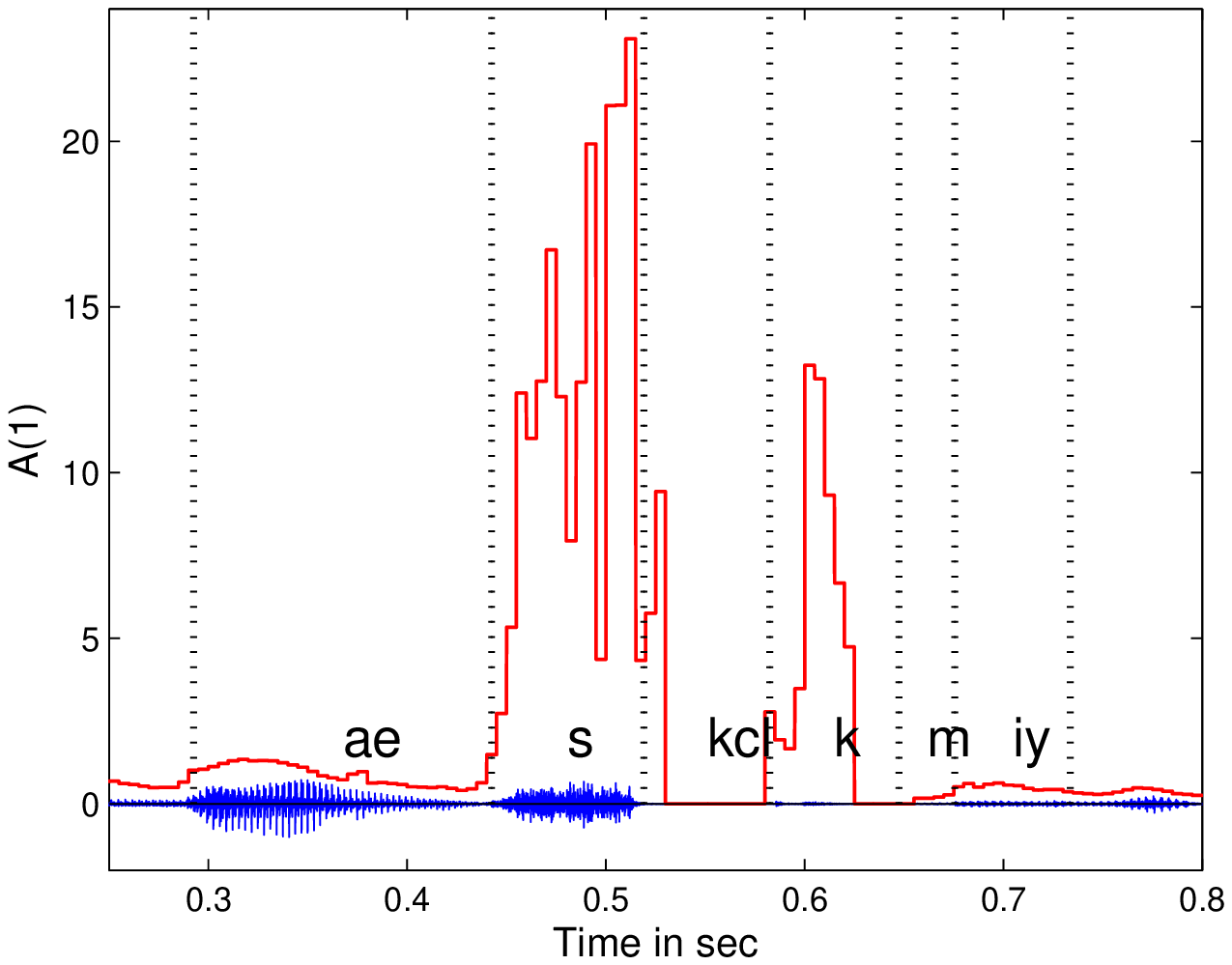}}
      \centering
      \subfigure[]{
      \includegraphics[width=3.5in,height=2in]{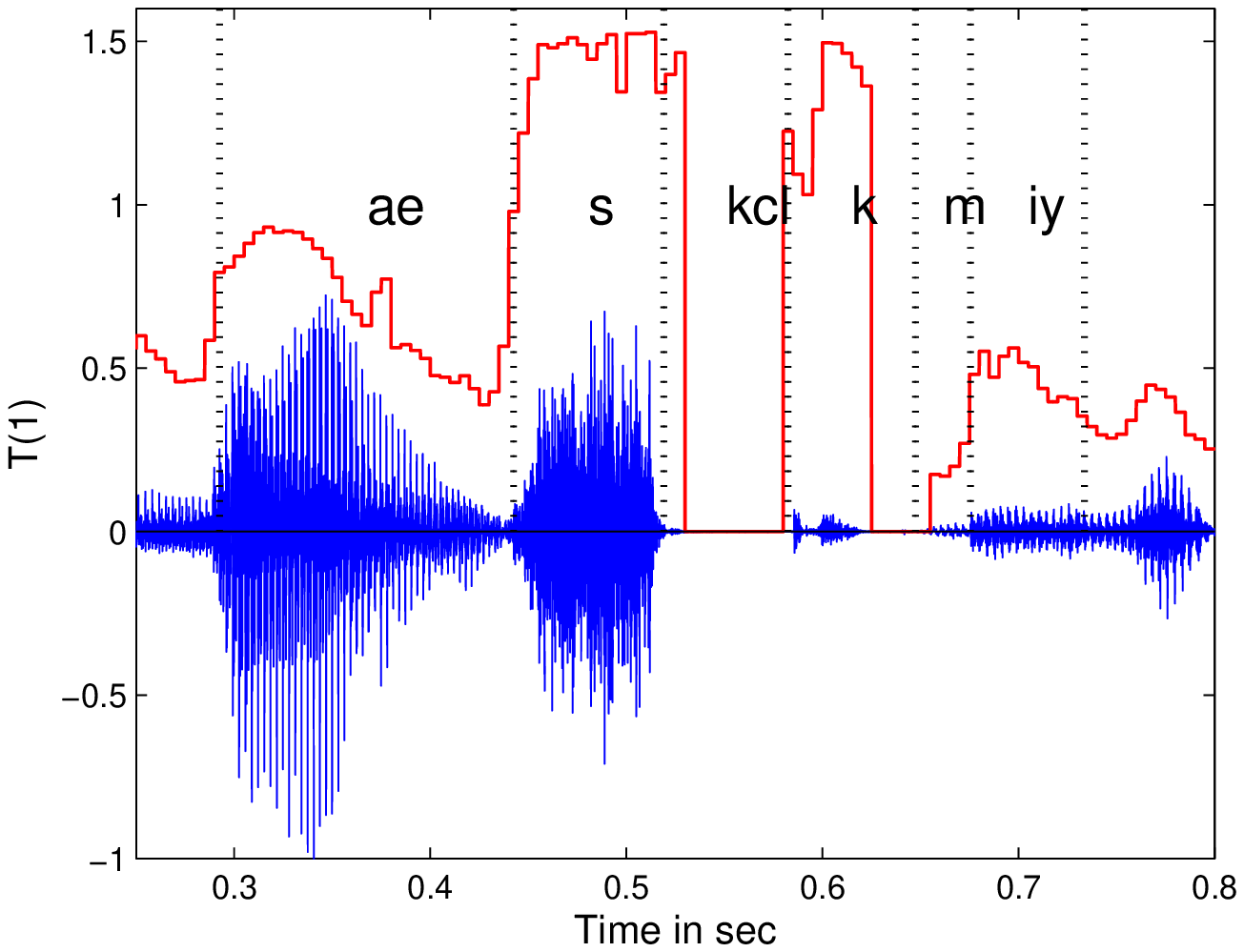}}}
      \caption{Segment of a speech signal comprising the utterance `ask me' (sonorant-fricative-stop-sonorant) and the plots of frame-wise values of (a) A(1) ; (b) T(1) in radians.}
      \end{figure}

\section{EXPERIMENTS AND RESULTS}
\label{s2}
 
\subsection{Histogram of SFDI for sonorants and fricatives}

The hand labeled TIMIT database \cite{23} is used for validating the proposed concept. The labeled boundaries in the TIMIT database are used to identify the sonorant and fricative segments. SFDI is computed frame-wise over these segments. For the purpose of computing histograms, `hh' is excluded from sonorants since it behaves like an unvoiced sound. Phones `th', `dh' and `v' are excluded from the fricative class, since these phones behave anamolously. The entire TIMIT database is analyzed. The value of SFDI is computed for all the frames within each segment of the two classes, sonorants and fricatives. The 
histograms of SFDI for the two classes, normalized by the number of occurrences of each class, are shown in Fig.2 for a bin size of 0.025. The two histograms show a clear separation with a very small overlap. The number of sonorants falls sharply for SFDI values above 1.1, whereas SFDI has values larger than 1.1 for the fricatives.

\begin{figure}
 \captionsetup{justification=raggedright,
      singlelinecheck=false
      }
\centering
\includegraphics[width=3.5in,,height=2in]{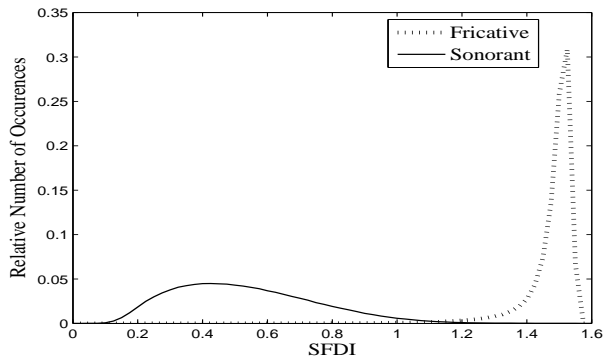}
\caption{Normalized histogram of the value of SFDI within sonorant (solid) and fricative (dashed) segments from the entire TIMIT database. }
 \label{(a)}
 \end{figure}
\vspace{-0.08cm}
\subsection{Arriving at the threshold and the frame-wise performance}

Assume that a threshold based logic is used and whenever SFDI is less than a threshold \textit{\textbf{T}}, the frame is assigned to the sonorant class; else, to the fricative class. If SFDI for a known frame of a sonorant (fricative) is lesser (greater) than the threshold \textit{\textbf{T}}, then that frame is considered to be correctly classified.  The ratio of the number of correctly classifed frames to the total number of frames studied gives the accuracy. The frame-wise accuracy is computed for the two classes on a development set of the TIMIT database, consisting of 1140 files. As the threshold is increased, the error rate falls sharply for the sonorants since lesser number of sonorant frames have a higher value of SFDI. On the other hand, as the threshold is increased, the area under the normalized histogram below the threshold increases thereby increasing the error rate for the fricatives. Figure 3 shows the frame-wise error rate Vs threshold. The cross-over point of error rates occurs at a threshold of 1.1 and the corresponding error rate is about 0.80\%. A similar method is used for arriving at  the threshold for the entire TIMIT database consisting of 6300 files. The threshold is found to be 1.1, which is the same as that obtained for the development set. Frame-wise error rate with this threshold is 0.93\%, which is about the same as that obtained for the development set. Some of the errors occur at the frame boundaries, where the hand labeled boundaries may not have been accurately marked. 

\begin{figure}
 \captionsetup{justification=raggedright,
      singlelinecheck=false
      }
\centering
\includegraphics[width=3.5in,,height=2in]{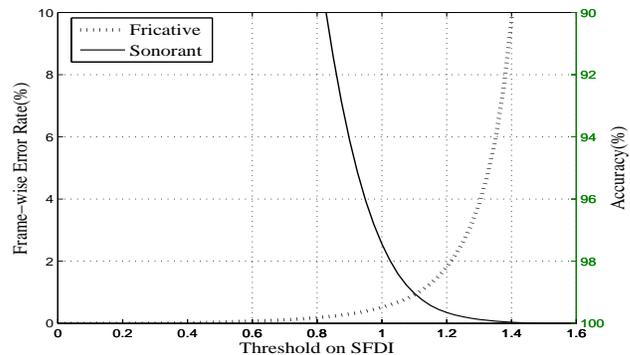}
\caption{Frame-wise error rate on the development set of TIMIT as a function of threshold on SFDI for sonorants (solid) and fricatives (dashed).}
 \label{(a)}
 \end{figure}

\begin{table}
\captionsetup{justification=raggedright,
      singlelinecheck=false
      }
      \caption{Effect of noise on the performance of SFDI on the development set and the entire database.\\
                            $^{**}$Orgnl: Original data set of TIMIT and NTIMIT, supposed to have SNR values of 39.5 dB and 26.5 dB, respectively. }
                     \centering
                     
                     \begin{tabular}{|p{0.8cm}|p{0.2cm}|p{0.4cm}|p{0.4cm}|p{0.2cm}|p{0.4cm}|p{0.4cm}|}
                     
                     \hline
                     \multirow{2}{*}{\textbf{}}
                      &  \multicolumn{3}{|c|}{\textbf{TIMIT}}&  \multicolumn{3}{|c|}{\textbf{NTIMIT}}\\
                        \cline{2-7}                    
                      
                      
                       \multicolumn{1}{|p{0.8cm}|}{\textbf{Noise Level}} 
                       & \multirow{2}{*}{\textbf{\textit{T}}}& \multicolumn{2}{|c|}{\textbf{Accuracy (\textbf{\%}})}& \multirow{2}{*}{\textbf{\textit{T}}}& \multicolumn{2}{|c|}{\textbf{Accuracy (\textbf{\%}})}\\

                    \cline{3-4} 
                    \cline{6-7}
                      \multicolumn{1}{|c|}{\textbf{}} & \multicolumn{1}{|c|}{\textbf{\textit{}}}& \multicolumn{1}{|c|}{\textbf{Dev. Set}}& \multicolumn{1}{|c|}{\textbf{Full Set}}& \multicolumn{1}{|c|}{\textbf{\textit{}}}& \multicolumn{1}{|c|}{\textbf{Dev. Set}}& \multicolumn{1}{|c|}{\textbf{Full Set}}\\
                     
                     \hline
                     \multicolumn{1}{|p{0.8cm}|}{Orgnl.$^{**}$} & \multicolumn{1}{|c|}{1.10}& \multicolumn{1}{|c|}{99.20}& \multicolumn{1}{|c|}{99.07}& \multicolumn{1}{|c|}{0.62}& \multicolumn{1}{|c|}{89.04}& \multicolumn{1}{|c|}{88.63}\\

                     \hline
                     \multicolumn{1}{|p{0.8cm}|}{20 dB} & \multicolumn{1}{|c|}{1.30}& \multicolumn{1}{|c|}{98.52}& \multicolumn{1}{|c|}{98.38}& \multicolumn{1}{|c|}{0.99}& \multicolumn{1}{|c|}{87.94}& \multicolumn{1}{|c|}{87.35}\\
                     
                     \hline
                      \multicolumn{1}{|p{0.8cm}|}{15 dB} & \multicolumn{1}{|c|}{1.36}& \multicolumn{1}{|c|}{98.21}& \multicolumn{1}{|c|}{98.04}& \multicolumn{1}{|c|}{1.12}& \multicolumn{1}{|c|}{87.10}& \multicolumn{1}{|c|}{86.59}\\
                      
                      \hline
                      \multicolumn{1}{|p{0.8cm}|}{10 dB} & \multicolumn{1}{|c|}{1.41}& \multicolumn{1}{|c|}{97.65}& \multicolumn{1}{|c|}{97.48}& \multicolumn{1}{|c|}{1.22}& \multicolumn{1}{|c|}{85.97}& \multicolumn{1}{|c|}{85.54}\\                      
                      
                      \hline
                      \multicolumn{1}{|p{0.8cm}|}{5 dB} & \multicolumn{1}{|c|}{1.43}& \multicolumn{1}{|c|}{95.78}& \multicolumn{1}{|c|}{95.65}& \multicolumn{1}{|c|}{1.29}& \multicolumn{1}{|c|}{84.05}& \multicolumn{1}{|c|}{83.91}\\     
                     
                      \hline
                      \multicolumn{1}{|p{0.8cm}|}{0 dB} & \multicolumn{1}{|c|}{1.44}& \multicolumn{1}{|c|}{91.53}& \multicolumn{1}{|c|}{91.35}& \multicolumn{1}{|c|}{1.33}& \multicolumn{1}{|c|}{80.83}& \multicolumn{1}{|c|}{80.44}\\                                       \hline    
                     \end{tabular} 
                       
                            \label{q1}
                      \end{table}

\subsection{Extended experiments}
It has been reported that the results obtained with a development set are comparable to that obtained with the entire database \cite{6,16,22}. Here, we present the results for both the development set as well as for the full set of TIMIT.

\begin{figure*}

\centering
\includegraphics[width=7.5in,height=2.4in]{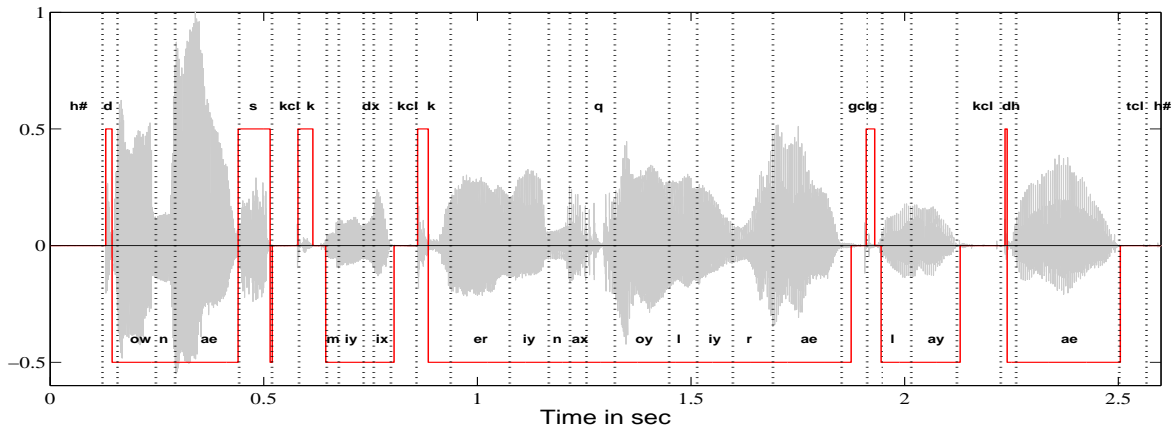}
\hspace{0.25cm}\caption{Two-class segmentation of the utterance `Don't ask me to carry an oily rag like that' using SFDI.}
 \label{(a)}
 \end{figure*}
\subsubsection{Robustness}
 In order to study the robustness of the measure SFDI, white Gaussian noise is added to the speech samples with an appropriate scale factor to achieve the desired global SNR. SNRs of 20 down to 0 dB, in steps of 5 dB, are considered in the experiment. The value of the threshold \textit{\textbf{T}} at the cross-over point is measured. The value of \textit{\textbf{T}} is about the same for both the development set and the entire database.  The corresponding frame-wise accuracy is shown in Table I.
                     
 As the SNR decreases, the threshold \textit{\textbf{T}} increases. Assuming the white noise and speech signal to be uncorrelated, the spectral level shifts uniformly across all frequencies and hence SFDI also increases. The frame-wise accuracy even at 0 dB SNR is 91.3\%, which is respectable.  

\subsubsection{Experiments on the NTIMIT database}  

The NTIMIT database is the telephone quality counterpart of the TIMIT database with a reduction in the bandwidth (300-3400 Hz) as well as a deterioration of SNR (39.5 to 26.8 dB)\cite{24}. Once again, the value of \textit{\textbf{T}} is noted to be the same for the development set and for the entire database The frame-wise accuracy for the original speech is around 89\%. The results for different SNRs are shown in Table I. The performance degrades with noise and the frame-wise accuracy is about 80.4\% at 0 dB SNR
\subsubsection{SFDI-based 2-class segmentation}
Using SFDI, speech signal is broadly segmented into two classes. The frames with SFDI $>\textbf{\textit{\textbf{T}}}$ are assigned to class-1 and others to class-2. A rectangular function is defined with values of 0.5 and -0.5 for the two classes. If the energy in a frame is lower than 0.0004 times the maximum frame energy, then the value of the rectangular function is forced to zero for that frame to signify a silence segment. The resultant three-level rectangular function is shown in Fig.4 along with the speech waveform for an utterance from TIMIT. The hand-labeled boundaries and corresponding labels are shown. Identified class-1 mostly comprises fricatives and class-2 mostly comprises sonorants and the detected boundaries lie close to the hand labeled boundaries between the two classes. It may be noted that for stops, the segments are very short in duration and are preceded by a silence segment (closure duration). Since the likely interval of the bursts can be deduced, the plosion index used for detecting a burst \cite{22} need to be computed only over such an interval instead of at every sample. This illustrates the potential use of SFDI for segmentation. Also, such a segmentation may be used for non-uniform frame rate analysis. A formal validation of segmentation, primarily using SFDI as a feature, is beyond the scope of the present paper.

\vspace{-0.08cm}
\subsection{Comparison with previous work}

Although a strict comparison with previous studies is not possible since the size of the database used and the tasks addressed are different, we make some broad observations for comparative purposes and these must not be construed as a criticism of the earlier results. 

For the manner classification task, the reported accuracy is in the range of 70-80\% \cite{6,9}. Frame-wise accuracy for the distinctive feature voiced/unvoiced is reported to be about 93\% and for vowel/fricative distinction, about 87\% for the TIMIT database and original speech [8]. The accuracy for [sonorant] or [fricative] detection is on an average about 95\% for TIMIT, using 42 acoustic parameters and SVM classifier \cite{7}. The highest reported accuracies for V/U classification for the TIMIT database are about 94.4\% for original speech and 92.7, 91.8, 89.7 and 86.4\% for  20,  10, 5  and  0 dB SNR, respectively for additive white noise \cite{17}. For the NTIMIT database, the reported classification accuracy is about 80\% based only on the algorithm, i.e., without correcting for the transcription errors \cite{16}. In comparison, the accuracy obtained in this study for a broad 2-class separation using a single scalar measure of SFDI is about 99.07\% for the original, 91.35\% at 0 dB SNR for the TIMIT database and 88.6\% for the original and 80.4\% at 0 dB SNR for the NTIMIT database. It has been observed that the results obtained for a development set and the entire database are about the same. 

\section{Conclusion}

This study has demonstrated that a simple scalar measure SFDI, viz., inverse tan of sum of LPCs, along with a threshold based logic, may be effectively used to distinguish the sonorants from the fricatives. The experiments show that the discrimination given by SFDI is high even at 0 dB SNR. The results obtained in this study are comparable, or in some respects, better than the state-of-the-art methods. 

Future research would be to utilize this property of SFDI, along with additional features, for automatic segmentation of a speech signal into different phonetic classes. It would be of interest to understand the reason for the observed property of SFDI by studying the relationship between the spectral envelope and the sum of LPCs.

\bibliographystyle{IEEEtran}
\bibliography{IEEEabrv,thes}
\end{document}